\documentclass[a4paper, twocolumn]{article}
\usepackage{graphicx}
\usepackage{amsmath}

\begin{document}

\title{Locally inequivalent four qubit hypergraph states}
\author{Xiao-yu Chen, Lei Wang \\
{\small {College of Information and Electronic Engineering, Zhejiang
Gongshang University, Hangzhou, 310018, China}}}
\date{}
\maketitle

\begin{abstract}
Hypergraph states as real equally weighted pure states are important
resources for quantum codes of non-local stabilizer. Using local Pauli
equivalence and permutational symmetry, we reduce the 32768 four qubit real
equally weighted pure states to 28 locally inequivalent hypergraph states
and several graph states. The calculation of geometric entanglement
supplemented with entanglement entropy confirms that further reduction is
impossible for true hypergraph states.

PACS number(s): 03.67.Mn, 03.65.Ud, 03.67.Ac

Keyword(s): hypergraph state; local equivalence; geometric
entanglement.
\end{abstract}

\section{Introduction}

A real equally weighted pure state (REW) is a superposition of all basis
states with real amplitudes and equal probabilities. Recently, it has been
systematically proven that REWs have a one-to-one correspondence with the
quantum hypergraph states or graph states\cite{Qu} \cite{Rossi}. Thus a REW
can be described by a mathematical hypergraph, namely, a graph where at
least one of its edges connecting more than two vertices; or a `usual' graph
with two vertex edges. A hypergraph state can be described by a non-local
stabilizer, in contrast to a usual graph state which is described by a local
stabilizer whose observables are simple tensor products of Pauli matrices.
There are a large number of hypergraph states even for a system of a few
qubits. Then the local equivalence of hypergraph states is an important
problem in applying hypergraph states for quantum information processing.
For the local equivalence of arbitrary multipartite pure quantum states, the
local polynomial invariants have been introduced \cite{Grassl} and the
necessary and sufficient conditions have been proposed \cite{Kraus}.
However, a more practical way is to use the entanglement to characterize the
local equivalence of states. In this paper, we will study the local
equivalence of all the four qubit hypergraph states and classify the states
with their geometric measure of entanglement supplemented with bipartite
entanglement entropy, both of them are easily calculable.

\section{Hypergraph states}

A hypergraph $H=(V;E)$ is composed of a set $V$ of $n$ vertices and a set of
hyperedges $E.$ The hyperedge set $E$ consists of $k-$hyperedges (hyperedge
connecting $k$ vertices) for $1\leq k\leq n.$ The rank of a hypergraph is
the maximum cardinality of its hyperedges. The $k-$hyperedge neighborhood $%
N_k(i)$ of the vertex $i$ is defined as $N_k(i)=\{\{i_1,i_2,...,i_{k-1}\}|%
\{i,i_1,i_2,...,i_{k-1}\}\in E\},$ where $\{i_1,i_2,...,i_{k-1}\}$ is the $%
k-1$ hyperedge. The neighborhood $N(i)$ of the vertex $i$ is $N(i)=\cup
_kN_k(i)$. We also use $N(i)$ to denote the corresponding set of neighbor
hyperedges. To associate the hypergraph state with the underlying
mathematical hypergraph, we assign each vertex a qubit and initialize each
qubit as the state $\left| +\right\rangle =\frac 1{\sqrt{2}}(\left|
0\right\rangle +\left| 1\right\rangle )$ ; each $k-$hyperedge represents the
$k-$body interaction among the corresponding $k$ qubits. The hypergraph
state related to the hypergraph $H$ is defined as
\begin{equation}
\left| H\right\rangle =\prod_{k=1}^n\prod_{\{i_1,i_2,\ldots ,i_k\}\in
E}U_{i_1i_2\ldots i_k}\left| +\right\rangle ^{\otimes n},
\end{equation}
where $U_{i_1i_2\ldots i_k}$ is the $k-$qubit control $Z$ gate such that $%
U_{i_1i_2\ldots i_k}\left| 11\ldots 1\right\rangle _{i_1i_2\ldots
i_k}=-\left| 11\ldots 1\right\rangle _{i_1i_2\ldots i_k}$ , and leaves all
the other components of the computational basis unchanged. The rank of a
hypergraph state is the rank of it corresponding hypergraph. We will refer
rank $3$ or higher hypergraph state as true hypergraph state (hereafter a
hypergraph state refers to a true hypergraph state). Hypergraph states can
also be put into stabilizer formalism \cite{Rossi} as graph states can. The
difference is that the stabilizer operators for a graph state are the
product of local Pauli operators, while the set of stabilizer operators for
a hypergraph state consists nonlocal control phase gate operators. Given a
general hypergraph, for any vertex $i$, the stabilizer operator is defined
as
\begin{equation}
K_i=X_i\prod_{k=1}^n\prod_{\{i_1,i_2,\ldots ,i_{k-1}\}\in
N_k(i)}U_{i_1i_2\ldots i_{k-1}},  \label{aee1}
\end{equation}
where $X_i$ is the Pauli $X$ operator (bit flip) of vertex $i$. The
hypergraph state $\left| H\right\rangle $ is stabilizes by an Abelian
stabilizer group with generator set $\{K_i\}$ \cite{Rossi}, namely,
\begin{equation}
K_i\left| H\right\rangle =\left| H\right\rangle .  \label{aee2}
\end{equation}

Considering the local equivalence (denoted as $\stackrel{Local}{=}$) of the
hypergraph states, we have $X_i\left| H\right\rangle \stackrel{Local}{=}%
\left| H\right\rangle $ since $X_i$ is a local operator. Using (\ref{aee1})
and (\ref{aee2}), we have
\begin{equation}
\prod_{k=1}^n\prod_{\{i_1,i_2,\ldots ,i_{k-1}\}\in N_k(i)}U_{i_1i_2\ldots
i_{k-1}}\left| H\right\rangle \stackrel{Local}{=}\left| H\right\rangle .
\end{equation}
Hence applying all the control $Z$ operation containing in the neighborhood
of a vertex to a hypergraph state, we obtain a locally equivalent hypergraph
state. In the underlying mathematical hypergraph, the transform of hyperedge
set produced by $X_i$ is \cite{Qu}
\begin{equation}
E\rightarrow E^{\prime }=N(i)\Delta E,  \label{aee3}
\end{equation}
where $\Delta $ is the symmetric difference, that is, $E\Delta F=E\cup
F-E\cap F.$ The local equivalence (\ref{aee3}) is a very useful tool in
classifying the hypergraphs. As an example, suppose there be a four vertex
hypergraph $H$ with $E=\{\{1,2,3,4\},\{1,2,3\}\}$, the local equivalent
hypergraph $H^{\prime }$ with $E^{\prime }=N(4)\Delta E$ can be deduced by
applying local operator $X_4$ of the fourth qubit to the hypergraph state.
Then $E^{\prime }=\{\{1,2,3,4\}\}.$ Hypergraph $H^{\prime }$ is shown as $%
No.1$ in Fig. 1. In fact we can remove all the $n-1$ hyperedges for an $n$
vertex hypergraph of rank $n$ by applying Pauli $X_i$ operators.

We also have $Z_i\left| H\right\rangle \stackrel{Local}{=}\left|
H\right\rangle $ for $Z_i$ is the Pauli $Z$ operator (phase flip) of vertex $%
i$. The transform of hyperedge set produced by $Z_i$ is \cite{Qu}
\begin{equation}
E\rightarrow E^{\prime }=\{\{i\}\}\Delta E,  \label{aee4}
\end{equation}
that is, a loop is added to (removed from) vertex $i$ when there isn't (is)
a loop. The unitary operator corresponds to a loop $\{\{i\}\}$ on vertex $i$
is $U_i=Z_i$ \cite{Rossi}. We can also define hypergraph basis states $%
\left| H_{\mathcal{C}}\right\rangle =Z^{\mathcal{C}}\left| H\right\rangle ,$
where $\mathcal{C}=(c_1,\ldots ,c_n)$ is a bit string with $c_i=0,1$, and $%
Z^{\mathcal{C}}=\otimes _{i=1}^nZ_i^{c_i}.$ Then all the local $Z$
equivalent hypergraphs can be written in the form of $H_{\mathcal{C}}.$

\section{Entanglement Measures}

The entanglement measures for a multipartite quantum pure states are the
Schmidt measure \cite{Hein}, the (logarithmic) geometric measure \cite{Wei},
the relative entropy of entanglement \cite{Vedral}, the logarithmic
robustness \cite{Vidal} and so on. The later three measures are equal for
graph states\cite{Hayashi}. Unfortunately, they are not equal for a (true)
hypergraph state. Geometric measure is the easiest one to be calculated
among these three entanglement measures for the multipartite entanglement.
An iterative algorithm was derived for the entanglement of a graph state
\cite{Chen}. The algorithm can also be applied to a hypergraph state after a
slight modification. Thus we will use geometric measure to study the
entanglement property of four qubit hypergraph states and classify the
states by their entanglement values.

For a four qubit hypergraph state, the iterative algorithm of the geometric
measure is as follows: let the hypergraph state be $\left| \psi
\right\rangle ,$ its closest product state be $\left| \Phi \right\rangle
=\prod_{i=1}^4\left| \phi _i\right\rangle ,$ with $\left| \phi
_i\right\rangle =x_i\left| 0\right\rangle +y_i\left| 1\right\rangle $ and $%
\left| x_i\right| ^2+\left| y_i\right| ^2=1.$ The overlap amplitude (inner
product) of the hypergraph state and its closest product state is $%
f=\left\langle \psi \right| \left. \Phi \right\rangle .$ The Lagrange
multiplier method of maximizing $\left| f\right| ^2$ subject to the
conditions $\left| x_i\right| ^2+\left| y_i\right| ^2=1$ ($i=1,\ldots ,4$)
gives the iterative equations
\begin{eqnarray}
x_i^{*} &=&\mathcal{N}_i\frac{\partial f}{\partial x_i}, \\
y_i^{*} &=&\mathcal{N}_i\frac{\partial f}{\partial y_i},
\end{eqnarray}
where $\mathcal{N}_i$ is the normalization. By solving the equations we
obtain the closest product state $\left| \Phi \right\rangle $. As far as $%
\left| \Phi \right\rangle $ is determined, it follows the overlap amplitude $%
f=\left\langle \psi \right| \left. \Phi \right\rangle $ of a give hypergraph
state $\left| \psi \right\rangle ,$ and the geometric measure of
entanglement of $\left| \psi \right\rangle $ is
\begin{equation}
E_g=-\log _2\left| f\right| ^2.
\end{equation}

Exact expression of the entanglement values for some of the hypergraph
states can also be obtained based on the numeric calculation of the closest
product states. As an example, let us consider the $No.12$ (in Fig. 2)
hypergraph state which is the direct product of $\left| +\right\rangle $
with
\begin{eqnarray}
\left| \psi _3\right\rangle  &=&\frac 1{\sqrt{8}}(\left| 000\right\rangle
+\left| 001\right\rangle +\left| 010\right\rangle +\left| 011\right\rangle
\nonumber \\
&&+\left| 100\right\rangle +\left| 101\right\rangle +\left| 110\right\rangle
-\left| 111\right\rangle ).  \label{waa}
\end{eqnarray}
The later is the hypergraph state of a three qubit hypergraph with a $3-$%
hyperedge and without further two vertex edges. The closest state of $\left|
\psi _3\right\rangle $ is assumed be $\left| \phi \right\rangle ^{\otimes 3}$
due to the symmetry of the three qubits, where $\left| \phi \right\rangle
=x\left| 0\right\rangle +y\left| 1\right\rangle $ with normalization $\left|
x\right| ^2+\left| y\right| ^2=1.$ Denote $z=y/x,$ then the iterative
equation is
\begin{equation}
z^{*}=(1+2z-z^2)/(1+z)^2.  \label{wee0}
\end{equation}
Numeric calculation shows that $z$ eventually converges to a real number in
a few steps regardless its randomly chosen initial complex value. Thus we
arrive at the algebraic equation $z^3+3z^2-z-1=0.$ The solution of which is $%
z=-1-\frac{4\sqrt{3}}3\cos (\tau +\frac{2\pi }3)$ $\approx 0.6751,$ where $%
\tau =\frac 13\arctan \sqrt{37/27}.$ The geometric measure of the state $%
\left| \psi _3\right\rangle $ is
\begin{equation}
E_g=-\log _2\left| \left\langle \psi _3\right. \left| \phi \right\rangle
^{\otimes 3}\right| ^2\approx 0.5647.  \label{wee1}
\end{equation}

The relative entropy of entanglement is $E_r=\min_\sigma -\left\langle \psi
_3\right| \log _2\sigma \left| \psi _3\right\rangle $ for a pure state $%
\left| \psi _3\right\rangle $ where $\sigma $ belongs to the fully separable
state set. However, the full separability for a generic three qubit system
is unknown. Hence the relative entropy of entanglement is not available. The
entanglement is lower bounded by the entanglement of a bipartition of the
hypergraph state, this is due to fact that the fully separable state set is
the subset of the biseparable state set. The minimization over a larger set
will give a lower value. The bipartite relative entropy of entanglement $%
E_{rbi}$ is simply the minimal entanglement entropy of all the bipartitions
for the pure symmetric state $\left| \psi _3\right\rangle $. We have $%
E_{rbi}=-Tr\rho \log _2\rho $ with $\rho =Tr_{23}\left| \psi _3\right\rangle
\left\langle \psi _3\right| =\frac 34\left| +\right\rangle \left\langle
+\right| +\frac 14\left| -\right\rangle \left\langle -\right| $ being the
reduced state by tracing the second and the third qubits, where $\left|
-\right\rangle =\frac 1{\sqrt{2}}(\left| 0\right\rangle -\left|
1\right\rangle )$. Thus
\begin{equation}
E_r\geq E_{rbi}\approx 0.8113.
\end{equation}
The relative entropy of entanglement is larger than the geometric measure
for the three qubit hypergraph state.

\section{Four qubit hypergraph states with a four vertex hyperedge\label%
{sec1}}

For the four qubit system, we have REW states
\begin{equation}
\left| \psi _{REW}\right\rangle =\frac 14\sum_\mu (-1)^{g(\mu )}\left| \mu
\right\rangle ,
\end{equation}
where $\mu $ is a four bit string and $g(\mu )$ is a Boolean function, i.e. $%
g:$ $\{0,1\}^{\otimes 4}\rightarrow \{0,1\},$ the coefficient $(-1)^{g(\mu )}
$ can be $\pm 1$ for each $\mu $. The state $\left| \psi _{REW}\right\rangle
$ is uniquely defined by the function $g$ via the signs (either plus or
minus) in front of each component $\mu $ of the computational basis. It is
clear that we have $2^{16}$ different expressions of the coefficient series.
Up to the overall phase, the total number of REW states is $2^{15}.$ There
is a one-to-one correspondence between an REW state and a hypergraph state
or graph state. The number of $i$ vertex hyperedges is $C_4^i=\frac{4!}{%
i!(4-i)!}$ . The number of four qubit hypergraph states with a four vertex
hyperedge (rank $4$) is $N_4=2^{14}$, which is a half of the the total
number of REW states. Applying the Pauli $Z_i$ operator to a hypergraph
state gives rise to a one-vertex hyperedge (namely, the loop) on the $ith$
vertex of the underlying hypergraph. We then remove all the loops of a
hypergraph by applying proper Pauli $Z$ operators to obtain the standard
hypergraph (hypergraph free of loops). On the other hand, all the three
vertex hyperedges within the four vertex hyperedge can be removed by
applying the Pauli $X_i$ operators to the hypergraph states. Hence we will
consider the standard hypergraphs with two vertex edges and an overall four
vertex hyperedge for the four qubit hypergraph states. The number of these
hypergraphs is $64$ up to local equivalence of the Pauli $X_i$ operators and
$Z_i$ operators. Permutational symmetry of the vertices leads to further
reduction of $64$ hypergraphs to $11$ hypergraphs as shown in Fig.1. The
geometric measure and bipartite entanglement are listed in Table I.

\begin{table}[tbp]
{\bfseries TABLE I.} The geometric measure of entanglement ($GE$) and
bipartite entanglement of four qubit rank 4 hypergraph states. Here $BE_2$
is the vector of bipartite entanglement of partitions $12|34$, $13|24$, $%
14|23$. $BE_1$ is the vector of bipartite entanglement of partitions $1|234$%
, $2|134$, $3|124$, $4|123$. $m$ is the degeneracy with respect to
permutational symmetry. The locations of qubits 1,\ldots,4 are shown as
No.12 in Fig.2. $a=0.6561$, $b=1.2624$, $c=1.6773$, $d=0.5436$, $e=0.9544$.
\\[0ex]
\par
\begin{center}
\begin{tabular}{lllll}
\hline\hline
No. & $m$ & $GE$ & $BE_2$ & $BE_1$ \\ \hline
1 & 1 & 0.3043 & a,a,a & d,d,d,d \\
2 & 6 & 0.8157 & a,b,b & e,e,d,d \\
3 & 3 & 1.4891 & a,c,c & e,e,e,e \\
4 & 12 & 0.8954 & b,b,b & e,e,d,e \\
5 & 12 & 1.5261 & b,c,c & e,e,e,e \\
6 & 4 & 0.8916 & b,b,b & e,e,e,e \\
7 & 4 & 1.1360 & b,b,b & e,e,d,e \\
8 & 3 & 1.1732 & c,b,c & e,e,e,e \\
9 & 12 & 1.4316 & b,c,c & e,e,e,e \\
10 & 6 & 1.1165 & c,b,c & e,e,e,e \\
11 & 1 & 1.1726 & b,b,b & e,e,e,e \\ \hline\hline
\end{tabular}
\par
\end{center}
\end{table}

\begin{figure}[tbp]
\includegraphics[ trim=-0.010000in 0.010000in 0.138042in 0.000000in,
height=2.5in, width=3.5in ]{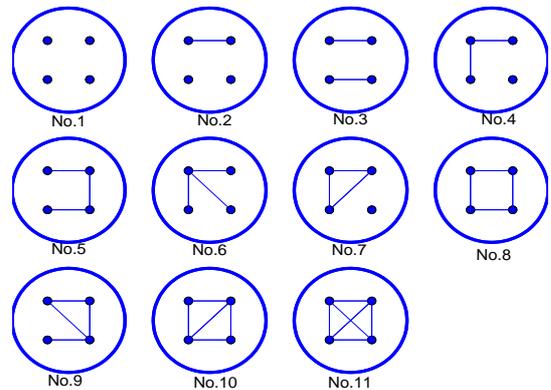}
\caption{(Color on line)The local inequivalent four vertex hypergraphs with
an overall four vertex hyperedge. The four vertex hyperedge is specified by
a circle, while the two vertex hyperedges are simply denoted by edges as in
\protect\cite{Rossi}. }
\end{figure}

The exact expression of the entanglement values for some hypergraph states
and further properties of the entanglement of hypergraph states with a four
qubit edge are shown in Table II.
\begin{table}[tbp]
{\bfseries TABLE II.} The properties of the four qubit rank 4 hypergraph
states. Here `PE' represents for the properties of the geometric
entanglement, it can be either exact or numerical. `D' represents for the
degeneracies of the closest product states. The closest product state types $%
|\phi_{1}\rangle^{4}$, $|\phi_{1}\rangle^{2}$$|\phi_{2}\rangle^{2}$, $%
|\phi_{1}\rangle$$|\phi_{2}\rangle^{2}$$|\phi_{3}\rangle$,$|\phi_{1}\rangle$$%
|\phi_{2}\rangle$$|\phi_{3}\rangle$$|\phi_{4}\rangle$ are denoted with `4',
`2,2',`1,2,1',`1,1,1,1', respectively. `R/C' mean the stable values of $%
z_{i} $ are real or imaginary. \\[0ex]
\par
\begin{center}
\begin{tabular}{llll}
\hline\hline
No. & PE & D & R/C \\ \hline
1 & Closed\text{ }form & 4 & R \\
2 & Numerical & 2,2 & R \\
3 & Numerical & 2,2 & R \\
4 & Numerical & 1,2,1 & R \\
5 & $3+2\log _2\frac 35 $ & 1,1,1,1 & R \\
6 & Numerical & 2,2 & R \\
7 & Numerical & 2,2 & C \\
8 & Closed\text{ }form & 4 & R \\
9 & Numerical & 1,2,1 & R \\
10 & Numerical & 2,2 & R \\
11 & $5-\log _2(9+3\sqrt{3}) $ & 4 & C \\ \hline\hline
\end{tabular}
\par
\end{center}
\end{table}

\section{Four qubit hypergraph states without a four vertex hyperedge \label%
{sec2}}

Unlike the four qubit hypergraph states with a four vertex hyperedge, the
hypergraph states without a four vertex edge have a different locally
equivalent classification with respect to Pauli $X$ operators. As can be
seen from $No.12$ hypergraph shown in Fig. 2, the locally equivalent states
are generated by $X_1,X_2$ and $X_3$ but not $X_4,$ where we denote the
vertex outside the ellipse as the fourth vertex, while the vertices inside
the ellipse are denoted as the first, second and third vertices,
respectively. The hypergraph state with underneath hypergraph $No.12$ is an
eigenstate of $X_4$. Thus the number of locally equivalent states produced
by Pauli $X$ operators is $8.$ This is also true for all the other
hypergraphs shown in Fig. 2. Together with the Pauli $Z$ local equivalence,
we have $128$ locally equivalent states for each hypergraph in Fig. 2. The
permutational symmetry of vertices gives rise to the degeneracy $m$ in Table
III. The total number of hypergraph states with three vertex hyperedges
(rank $3$) are $N_3=128\times \sum m=128\times 120=15\times 2^{10}.$ The
rank $2$ hypergraphs are just graphs. For graph states, the Pauli $X$
operators are equivalent to some other Pauli $Z$ operators and their tensor
products when considering local equivalence. Up to local Pauli $Z$
equivalence, the number of the graphs is $64$ which is the number $\sum m$
in Table. I. The total number of graphs is $N_2+N_1+N_0=16\times 64=2^{10}.$
Hence the number of hypergraph states of rank 3 together with graph states
is $N_3+N_2+N_1+N_0=2^{14}.$

The geometric measure and bipartite entanglement are listed in Table III.

\begin{table}[tbp]
{\bfseries TABLE III.} The geometric measure of entanglement and bipartite
entanglement of four qubit rank 3 hypergraph states. Here $BE_2$ is vector
of the bipartite entanglement of partitions $12|34, 13|24, 14|23$. $BE_1$ is
vector of the bipartite entanglement of partitions $1|234, 2|134$, $3|124,
4|123$. The locations of qubits 1,\ldots,4 are shown as No. 12 in Fig.2. $%
r=0.8113,$ $s=1.5,$ $t=1.2238,$ $u=1.6009.$ \\[0ex]
\par
\begin{center}
\begin{tabular}{lllll}
\hline\hline
No. & $m$ & $GE$ & $BE_{2}$ & $BE_{1}$ \\ \hline
12 & 4 & 0.5647 & r,r,r & r,r,r,0 \\
13 & 12 & 1.5417 & s,s,s & 1,r,1,1 \\
14 & 12 & 1 & s,s,r & 1,r,r,1 \\
15 & 4 & 1.5261 & s,s,s & 1,1,1,1 \\
16 & 6 & 0.6115 & r,t,t & r,r,r,r \\
17 & 6 & 1.2284 & r,u,u & 1,1,r,r \\
18 & 12 & 1 & s,t,t & r,r,r,1 \\
19 & 12 & 1.4150 & s,u,u & 1,1,r,1 \\
20 & 6 & 1.4569 & s,t,t & r,r,1,1 \\
21 & 6 & 1.4569 & s,u,u & 1,1,1,1 \\
22 & 4 & 1 & t,t,t & 1,r,r,r \\
23 & 12 & 0.6781 & t,t,t & r,r,r,r \\
24 & 12 & 1.3173 & u,u,t & 1,1,1,r \\
25 & 4 & 1.4150 & u,u,t & r,1,1,r \\
26 & 1 & 1.2230 & t,t,t & 1,1,1,1 \\
27 & 6 & 1.2767 & t,u,u & r,r,1,1 \\
28 & 1 & 0.8301 & t,t,t & r,r,r,r \\ \hline\hline
\end{tabular}
\par
\end{center}
\end{table}

\begin{figure}[tbp]
\includegraphics[ trim=1.30000in 0.000000in 0.0in 0.000000in, height=4.5in,
width=5in ]{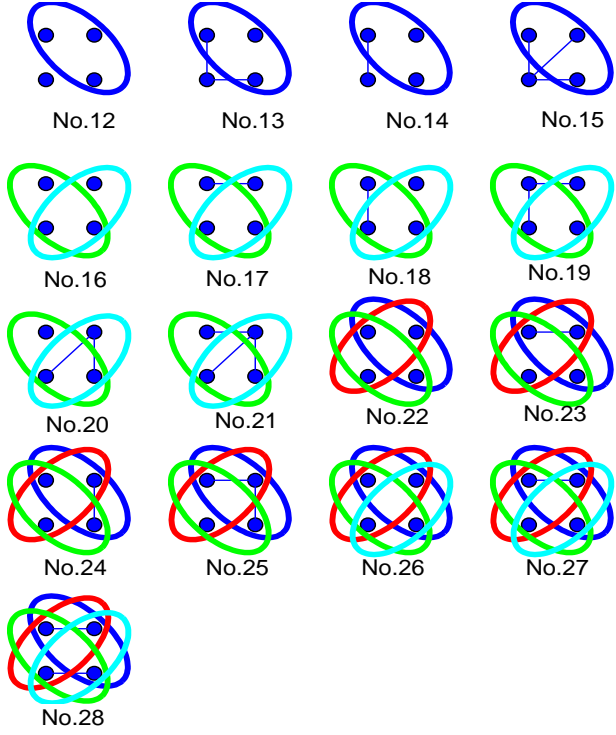}
\caption{(Color on line) The local inequivalent four vertex hypergraphs
without an overall four vertex hyperedge. The three vertex hyperedges are
specified by closed curves, while the two vertex hyperedges are simply
denoted by edges as in \protect\cite{Rossi}.}
\end{figure}

The exact expression of the entanglement values for some hypergraph states
and further properties of the entanglement of hypergraph states with three
qubit edges are shown in Table IV.

\begin{table}[tbp]
{\bfseries TABLE IV.} The properties of four qubit rank 3 hypergraph states.
Here `PE' represents for the properties of the geometric entanglement, it
can be either exact or numerical. `D' represents for the degeneracies of the
closest product states. The closest product state types $|\phi_{1}%
\rangle^{4} $, $|\phi_{1}\rangle$$|\phi_{2}\rangle^{3}$,$|\phi_{1}%
\rangle^{2} $$|\phi_{2}\rangle^{2}$, $|\phi_{1}\rangle$$|\phi_{2}\rangle^{2}$%
$|\phi_{3}\rangle$,$|\phi_{1}\rangle$$|\phi_{2}\rangle$$|\phi_{3}\rangle$$%
|\phi_{4}\rangle$ are denoted with `4', `1,3', `2,2',`1,2,1',`1,1,1,1',
respectively. `R/C' mean the stable values of $z_{i} $ are real or
imaginary. \\[0ex]
\par
\begin{center}
\begin{tabular}{llll}
\hline\hline
No. & PE & D & R/C \\ \hline
12 & Closed\text{ }form & 1,3 & R \\
13 & Numerical & 1,2,1 & R \\
14 & 1 & 1,3 & R \\
15 & $3+2\log _2\frac 35 $ & 1,3 & R \\
16 & $4-2\log _2(1+\sqrt{5})$ & 2,2 & R \\
17 & $2.5-\log _2(1+\sqrt{2})$ & 2,2 & C \\
18 & 1 & 1,3 & R \\
19 & $3-\log _23$ & 1,2,1 & R \\
20 & $4-2\log _2(1+\sqrt{2})$ & 1,2,1 & R \\
21 & $4-2\log _2(1+\sqrt{2})$ & 2,2 & R \\
22 & 1 & 1,3 & R \\
23 & $3-\log _25$ & 1,3 & R \\
24 & Numerical & 1,2,1 & R \\
25 & $3-\log _23$ & 1,2,1 & R \\
26 & $6-2\log _2(3+\sqrt{5})$ & 4 & R \\
27 & Numerical & 2,2 & R \\
28 & $4-2\log _23$ & 4 & R \\ \hline\hline
\end{tabular}
\par
\end{center}
\end{table}

\section{Discussion and Conclusion}

Each of the rank $4$ hypergraphs in section \ref{sec1} has a characteristic
value of geometric entanglement. Hence these $11$ hypergraph states are
confirmed to be all locally inequivalent due to their different values of
geometric entanglement. The entanglement entropy alone is not a good
indication for local inequivalent even considering all possible
bipartitions. We can see that $No.4$ and $No.7$ hypergraph states are not
discriminated by the series of bipartite entanglement entropy. The spectra
of bipartite entanglement entropy for $No.5,$ $No.8,$ $No.9$ and $No.10$
hypergraph states are identical or identical under the permutation of
qubits. The geometric entanglement of $No.5$ hypergraph state coincides with
that of $No.15,$ however, they are discriminated by their entanglement
entropy. The rank $3$ hypergraphs in section \ref{sec2} are all different by
their geometric entanglement values except for three cases where further
consideration of entanglement entropy is necessary. The three cases are that
the values of geometric entanglement of $No.19$ and $No.25$ hypergraph
states are equal; the values of geometric entanglement of $No.20$ and $No.21$
hypergraph states are equal; the values of geometric entanglement of $No.14,$
$No.18$ and $No.22$ hypergraph states are also equal. Further
discriminations are carried out by their bipartition entanglement entropy.

We conclude that there are $28$ locally inequivalent four qubit hypergraph
states, $11$ of them are rank $4,$ and $17$ of the them are rank $3.$ They
can be discriminated by geometric entanglement supplemented with bipartition
entanglement entropy. Local Pauli equivalence together with permutation
symmetry are enough in discriminating inequivalent four qubit hypergraph
states (it may not be true for hypergraph states with more qubits). Further
local Clifford equivalence such as local complimentary transform is not
necessary (it is also limited to four qubit hypergraph states), although it
is a powerful tool in discriminating local inequivalent graph states. We
have given a complete classification of the real equally weighted four qubit
pure states.

Note added: After submission of this work, we became aware of a recent
preprint by O. Guhne \textit{et al} \cite{Guhne}, which shows a similar
result.

\section*{Acknowledgement}

We thank the National Natural Science Foundation of China (Grant Nos.
11375152, 60972071) for support.

\end{document}